\documentclass[aps,pra,showpacs,twocolumn,amsmath,amssymb]{revtex4-1}
\usepackage{graphicx}% Include figure files
\usepackage{dcolumn}% Align table columns on decimal point
\usepackage{bm}% bold math
\usepackage{amsmath}
\usepackage{amssymb}
\usepackage{latexsym}
\usepackage{epsfig}
\usepackage{amsbsy}
\usepackage{array}
\usepackage{amssymb}
\usepackage{setspace}
\usepackage{bm}

\begin{document}
%%%%%%%%%%%

\title{Improving Cooling performance of the mechanical resonator with the two-level-system defects}
%\date{\today}
\author{Tian Chen$^{1,2}$}

\author{Xiang-Bin Wang$^{1,2,3}$}\email{xbwang@mail.tsinghua.edu.cn}

\affiliation{
$^1$State Key Laboratory of Low
Dimensional Quantum Physics, Department of Physics, Tsinghua University, Beijing 100084,
People's Republic of China\\
$^2$Synergetic Innovation Center of Quantum Information and Quantum Physics, University of Science and Technology of China, Hefei, Anhui 230026, People's Republic of China\\
$^3$Jinan Institute of Quantum Technology, Shandong
Academy of Information and Communication Technology, Jinan 250101,
People's Republic of China
}
\begin{abstract}
We study cooling performance of a realistic mechanical resonator containing defects. The normal cooling method through an optomechanical system does not work efficiently due to those defects. We show by employing periodical $\sigma_z$ pulses, we can eliminate the interaction between defects and their surrounded heat baths up to the first order of time. Compared with the cooling performance of no $\sigma_z$ pulses case, much better cooling results are obtained. Moreover, this pulse sequence has an ability to improve the cooling performance of the resonator with different defects energy gaps and different defects damping rates.
\end{abstract}


\maketitle

\section{introduction}
Preparing a ground state of a mechanical resonator is an important topic, and it has applications in testing fundamental quantum theory, exploring the boundary between classical and quantum regions, and studying precision metrology~\cite{Aspelmeyer_Rev13, Teufel_nature11, Verhagen_nature12}. Considering the inevitable interaction between the resonator and its surrounded heat bath, the state of the resonator is far away from the ground state in equilibrium, so cooling the mechanical resonator to the ground state becomes an urgent task. So far, many proposals have been put forward to cool the resonator. Making use of the capacitive coupling, Lorentz force, magnetic field or strain field induced coupling with a two-level-system (Josephson qubit, or negatively charged nitrogen-vacancy center, etc.), one can cool the mechanical resonator efficiently~\cite{Martin_PRB04, Wilson-Rae_PRL04, Zhang_PRL05, Jaehne_NJP08, Wang_NJP08, Rabl_PRB09, Zippilli_PRL09, Rabl_PRB10, Treutlein_arXiv12, Kepesidis_PRB13}. Other proposals design an optomechanical system consisting of one cavity mode and the resonator to be cooled. The radiative pressure from photons is applied to cool the resonator~\cite{Law_PRA95, Marquardt_PRL07, Wilson-Rae_PRL07, Wilson-Rae_NJP08, Genes_PRA08, Liu_PRL13}. For all of these proposals, we can finally obtain a resonator with very small phonon number in the long time limit.

In a realistic experimental optomechanical system, the mechanical resonator is often made by silica. Because of the amorphous nature of silicon, defects reside in the amorphous native oxide of the silicon surface~\cite{Anderson_72, Phillips_87, Enss_05, Ramos_PRL13}. As shown in Ref.~\cite{Tian_PRB11}, due to the coupling between the defect with resonator and their couplings with heat baths, the thermal noise from the heat bath of the defects can be effectively transferred to the mechanical resonator. Therefore, the normal cooling method used in the optomechanical system can actually hardly work.

 %and for simplicity, these defects can be seen as many two-level-systems. Considering the existence of a strain field in this material, the mechanical resonator can couple to the defects. As demonstrated that, from the optomechanical output spectrum, the signature of the defects can be seen clearly~\cite{Ramos_PRL13}. Moreover, when discussing cooling performance of the mechanical resonator in such an optomechanical system, , and the resonator is heated instead~\cite{Tian_PRB11}. So in the study of cooling the mechanical resonator, how to eliminate the bad effect from the defects is worthwhile to research.

In this paper, we show by employing the periodical $\sigma_z$ pulses, we can efficiently remove the detrimental effect of defects and cool the resonator efficiently. The reason is that the periodical $\sigma_z$ pulses can induce the sign of operators $\sigma_-$, $\sigma_+$ to be flipped~\cite{Viola_PRL99,Khodjasteh_PRL05, Khodjasteh_PRA07, Uhrig_PRL07, Yang_PRL08, Uhrig_PRL09, West_PRL10}. The interaction between defects and the surrounded heat bath can be eliminated up to the first order of time. We display the cooling performance of the resonator with different defects energy gaps and defects damping rates. We find that with a large number of $\sigma_z$ pulses ($N=99$), the phonon occupation of the resonator reduces to a lower value, when compared with the case of no $\sigma_z$ pulses. Besides, we also study the difference of the cooling results through two different calculation approaches, one uses the master equation based on polariton doublets, the other is "simple approach" in which the defects and coupling are added into the master equation of the bare resonator directly~\cite{Tian_PRB11}. In our discussion, we find that these two approaches give a similar qualitative picture in cooling, but the obtained value of phonon occupations are quantitative different.

The structure of this paper is as follows, Sec.~II introduces the total system we shall study. In Sec.~III, we explore the cooling performance with different defects energy gaps and different defects damping rates. The paper is concluded by Sec.~IV.

\section{Model}
The total system contains one cavity mode, a mechanical resonator, defects and their surrounded heat baths. The total Hamiltonian is~\cite{Ramos_PRL13, Tian_PRB11},
\begin{equation}
H_{tot}=H_{om}+H_{JC}+H_{a,e}+H_{b,e}+H_{\sigma,e}+H_{B}.
\end{equation}
Here, $H_{B}$ denotes the Hamiltonian of the three non-interacting baths. An optomechanical component consists of one cavity mode and the mechanical resonator, that is,
\begin{equation}
H_{om}=-\hbar\Delta_La^\dagger a+\hbar g(a+a^\dagger)(b+b^\dagger),
\end{equation}
where $\Delta_L=\omega_L-\omega_c$ is the detuning of cavity driving frequency $\omega_L$ from the cavity mode frequency $\omega_c$. The operators $a$ ($a^\dagger$) and $b$ ($b^\dagger$) stand for the annihilation (creation) operator of the cavity mode and mechanical resonator, respectively.

As discussed above, the defects couple to the mechanical resonator inevitably. Here, we only consider the case of one defect, which can be regarded as a two-levle-system (TLS)~\cite{Anderson_72, Phillips_87, Enss_05, Ramos_PRL13, Tian_PRB11}. The Hamiltonian for the resonator and the defect can be written as a Jaynes-Cummings (JC) form ~\cite{Ramos_PRL13, Tian_PRB11},
\begin{equation}
H_{JC}=\hbar\omega_mb^\dagger b+\frac{1}{2}\hbar\omega_z\sigma_z+\hbar\lambda(\sigma_+ b+b^\dagger\sigma_-),
\end{equation}
where $\omega_z(\omega_m)$ is the frequency of a TLS (mechanical resonator), and $\lambda$ is the coupling strength.

In our discussion below, we take the interactions between each of these three systems with their surrounding heat bath into account. The interaction Hamiltonian takes the form as,
\begin{subequations}
\begin{align}
&H_{a,e}=\sum_{k}g_{a,k}(aa^{\dagger}_{k}+a^{\dagger}a_{k}),\\
&H_{b,e}=\sum_{k}g_{b,k}(bb^{\dagger}_{k}+b^{\dagger}b_{k}),\\
&H_{\sigma,e}=\sum_{k}g_{\sigma,k}(\sigma_-c^{\dagger}_k+\sigma_+ c_{k}).
\end{align}
\end{subequations}
Here, $g_{a,k}$, $g_{b,k}$, and $g_{\sigma,k}$ are the coupling strength between the cavity mode, the mechanical resonator, the defects and each one's heat bath. The bath modes are labeled by $k$.

 %denote the coupling strength between one cavity mode and the heat bath, between the mechanical resonator and the heat bath, and between the defect and the heat bath, respectively. $k$ labels the number of bath modes.

In many cases, the coupling strengths $g$ and $\lambda$ are comparable, and the cavity damping rate is large. We define polariton states as~\cite{Tian_PRB11},
\begin{equation}
|n,\alpha\rangle=c_{\alpha}^{n}|n\downarrow\rangle+s_{\alpha}^n|(n-1)\uparrow\rangle,
\end{equation}
with $n\geq1$ and $\alpha=\pm$, $|n\rangle$ is the Fock state of the mechanical resonator, $|\uparrow\rangle$ and $|\downarrow\rangle$ are the eigenstates of $\sigma_z$. Here, $c_+^n=-s_-^n=\cos(\delta_n/2)$, $s_+^n=c_-^n=\sin(\delta_n/2)$, the expression of $\delta_n$ satisfies the relation, $\cos(\delta_n/2)=\sqrt{(\omega_n+\delta\omega)/2\omega_n}$, where $\delta\omega=\omega_m-\omega_z$, and $\omega_n=\sqrt{\delta\omega^2+4\lambda^2n}$.

By employing the projection operator technique~\cite{Cirac_PRA92, Breuer_02}, and only keeping terms up to the second order of $g$, we can obtain the dynamics of the reduced density matrix $\rho_s$ for the composite system of the TLS and resonator~\cite{Tian_PRB11},
\begin{equation}
\begin{split}
\dot{\rho}_s=&-\frac{i}{\hbar}[H_\tau,\rho_s]+\sum_{n,\alpha,\beta}\frac{\Gamma_0^{n\alpha\beta}}{2}\mathcal{L}_0^{n\alpha\beta}\rho_s\\&+\sum_{n,\alpha,\beta}
|A_{\beta,\alpha}^{(n)}|^2[\frac{\Gamma^n_{-,\alpha\beta}}{2}\mathcal{L}(O_n^{\alpha\beta})+\frac{\Gamma_{+,\alpha\beta}^{n}}{2}\mathcal{L}(O_{n}^
{\alpha\beta\dagger})]\rho_s,\label{tian}
\end{split}
\end{equation}
here, $\mathcal{L}_0^{n\alpha\beta}=(n_{th}^{n\alpha\beta}+1)\mathcal{L}(O_{n}^{\alpha\beta})+n_{th}^{n\alpha\beta}\mathcal{L}(O_{n}^{\alpha\beta\dagger})$, and $\mathcal{L}(o)\rho=2o\rho o^\dagger-\rho o^\dagger o-o^\dagger o\rho$. The polariton Hamiltonian $H_\tau=\omega_{n,\alpha}|n,\alpha\rangle\langle n,\alpha|$, with $\omega_{n,\alpha}$ is the eigenenergies of the polariton states. The thermal population $n_{th}^{n\alpha\beta}=(\exp(\hbar\omega_{n\alpha\beta}/k_BT)-1)^{-1}$, with $\omega_{n\alpha\beta}=\omega_{n,\alpha}-\omega_{n-1,\beta}$. Here, all heat baths have the same temperature $T$. The operator $O_n^{\alpha\beta}=|(n-1),\beta\rangle\langle n,\alpha|$. The expressions of $\Gamma_0^{n\alpha\beta}$ and $\Gamma_{\mp,\alpha\beta}^{n}$ are,
\begin{subequations}
\begin{align}
&\Gamma_0^{n\alpha\beta}=|A_{\beta,\alpha}^{(n)}|^2\gamma_m+|\sigma_{\beta,\alpha}^{(n)}|^2\gamma_\tau,\\
&\Gamma_{\mp,\alpha\beta}^{n}=\frac{g^2\kappa}{\kappa^2/4+(\omega_{n\alpha\beta}\pm\Delta_b)^2},
\end{align}
\end{subequations}
where $\gamma_m$, $\gamma_\tau$ and $\kappa$ stand for the damping rate of the mechanical resonator, the TLS, and the cavity mode, respectively. The coefficients $A_{\beta,\alpha}^{(n)}$ and $\sigma_{\beta,\alpha}^{(n)}$ are from the expressions of $b=\sum A_{\beta,\alpha}^{(n)}O_n^{\alpha\beta}$, $\sigma_-=\sum \sigma_{\beta,\alpha}^{(n)}O_n^{\alpha\beta}$.

\textit{$\sigma_z$ pulses}. Many researches have been devoted to designing different pulse sequences, to keep the system away from decoherence induced by the surrounded heat bath~\cite{Viola_PRL99,Khodjasteh_PRL05, Khodjasteh_PRA07, Uhrig_PRL07, Yang_PRL08, Uhrig_PRL09, West_PRL10}. Here, we use the periodical $\sigma_z$ pulses, to eliminate the interaction between the TLS and the bath up to the first order of time. Considering the properties of the operator $\sigma_z$,
\begin{equation}
\sigma_z \sigma_- \sigma_z=-\sigma_-,\quad \sigma_z \sigma_+ \sigma_z=-\sigma_+,
\end{equation}
we can obtain,
\begin{equation}
\begin{split}
&\sigma_z e^{-iH_{tot}t}\sigma_z e^{-iH_{tot}t}\\=&e^{-it\sigma_zH_{tot}\sigma_z}e^{-iH_{tot}t}
=e^{-itf_1+itf_2}\cdot e^{-itf_1-itf_2},\label{sigmaz}
\end{split}
\end{equation}
with the coefficients $f_1=H_{om}+\hbar\omega_mb^\dagger b+\frac{1}{2}\hbar\omega_z\sigma_z+H_{a,e}+H_{b,e}+H_B$, $f_2=\hbar\lambda(\sigma_+ b+b^\dagger\sigma_-)+H_{\sigma,e}$.
Clearly, up to the first order of time, we can eliminate the interaction of the TLS and the corresponding heat bath. Combing Eq.~(\ref{tian}) and (\ref{sigmaz}), we obtain the time evolution for the density matrix of $\rho_s(t)$.

%As pointed out in Ref.~\cite{Tian_PRB11}, this interaction makes the cooling performance of the mechanical resonator bad. Following Eq.~(\ref{tian}), we can obtain the dynamics of the reduced density matrix of TLS-resonator coupled system $\rho_s$ incorporating a periodical $\sigma_z$-pulses sequence as,
\begin{equation}
\begin{split}
&\rho_s(t)\Rightarrow\\ &\cdots e^{\mathcal{H}(t_j-t_{j-1})}\sigma_z\cdots\sigma_z \{e^{\mathcal{H}(t_2-t_1)}\sigma_z \{e^{\mathcal{H}t_1}\rho_{s}\}\sigma_z\}\sigma_z\cdots\sigma_z\cdots.
\end{split}
\end{equation}
Here, $e^{\mathcal{H}(t_{k}-t_{k-1})}\rho_{s}$ denotes the system evolution obeying Eq.~(\ref{tian}), $t_{k}-t_{k-1}$ is the time duration between two adjacent pulses. Within these, we can now investigate the cooling performance of the resonator numerically.
%Based on the discussion above, we think that we can improve the cooling performance of resonator by injecting the periodical $\sigma_z$ pulses.

\section{Results and Discussions}
\begin{figure}[htbp]
\begin{center}
\includegraphics[width=0.5\textwidth]{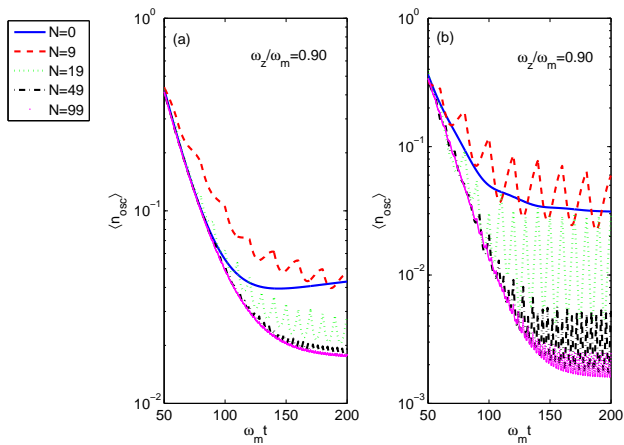}
\end{center}
\caption{\label{fig1}(Color online): Residual phonon number ($\langle n_{osc}\rangle$) versus different number of $\sigma_z$ pulses. (a) master equation based on polariton doublets. (b) "simple approach". Parameters set: $\omega_z/\omega_m=0.9$, $\kappa/\omega_m=0.15$, $\gamma_m/\omega_m=10^{-6}$, $\gamma_\tau/\omega_m=2.5*10^{-4}$, $g/\omega_m=0.05$, $\lambda/\omega_m=0.05$, $\Delta_L/\omega_m=-1$, $T=0.1K$.}
\end{figure}

Firstly, by using the master equation based on polariton doublets, we discuss the cooling performance of the mechanical resonator with different number of $\sigma_z$ pulses , see Fig.~\ref{fig1}.~(a). The frequency of the mechanical resonator is set as $\omega_m=200$ MHz. As a result, the more $\sigma_z$ pulses are used, the lower the resonator phonon number is.
%Here, we discuss cooling results with different incident pulses numbers. The phonon number of the resonator can be lowered to the smallest value when the most incident pulses are injected ($N=99$). .
If the number of $\sigma_z$ pulses is too small ($N=9$), the result is even worse than the case of no $\sigma_z$ pulses ($N=0$). Although based on the analysis above, the periodical $\sigma_z$ pulses can eliminate the interaction between TLS and its surrounding bath up to the first order of time, when the time interval between two adjacent $\sigma_z$ pulses is too long, the consequence of higher order terms are significant. When there are more $\sigma_z$ pulses, the time interval between two adjacent incident pulses is smaller, and the effect of those higher order terms becomes less significant. As shown in Fig.~\ref{fig1}.~(a), we achieve good cooling performance with the pulse numbers $N=19$, 49, 99. For the case of no $\sigma_z$ pulses, at time $\omega_mt=200$, the residual phonon number of the resonator is 0.04289, while for the case of $N=99$, the residual phonon number of the resonator is 0.01793 at time $\omega_mt=200$. There is a decrease of 58.2\%. In Fig.~\ref{fig1}.~(b), the cooling results by using the "simple approach" is presented. Same initial states are chosen for both two approaches~(Fig.~\ref{fig1}.~(a) and (b)). We find that the evolution of the population of the resonator given by "simple approach" is qualitatively same as the case shown in Fig.~\ref{fig1}.~(a), but the residual phonon number of the resonator in the long-time limit through these two approaches are quantitatively different. As pointed out in Ref.~\cite{Tian_PRB11}, results from the master equation based on polariton doublets (Fig.~\ref{fig1}.~(a)) are more accurate. From Fig.~\ref{fig1}.~(a), it is clearly seen that we can cool the resonator more efficiently with more injecting pulse number.

\begin{figure}[htbp]
\begin{center}
\includegraphics[width=0.5\textwidth]{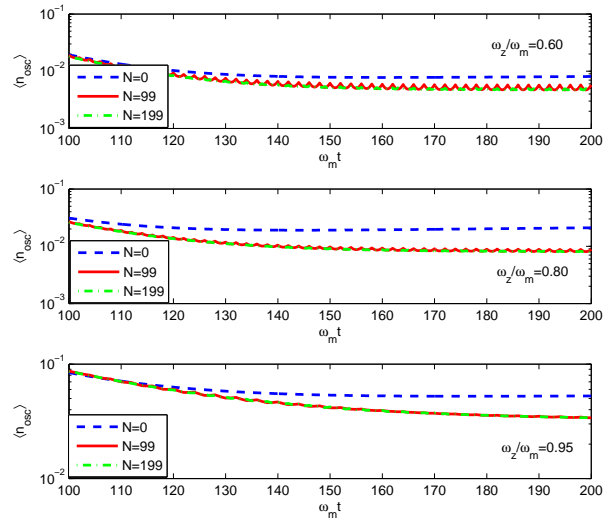}
\end{center}
\caption{\label{fig2}(Color online): Residual phonon number ($\langle n_{osc}\rangle$) versus different number of $\sigma_z$ pulses ($N$=0, 99, 199). Parameters set: $\kappa/\omega_m=0.15$, $\gamma_m/\omega_m=10^{-6}$, $\gamma_\tau/\omega_m=2.5*10^{-4}$, $g/\omega_m=0.05$, $\lambda/\omega_m=0.05$, $\Delta_L/\omega_m=-1$, $T=0.1K$. From top to bottom, $\omega_z/\omega_m$ is 0.6, 0.8, and 0.95.}
\end{figure}

Secondly, we study the cooling performance of the resonator with different defect energy gaps. Fig.~\ref{fig2} presents the results of three different cases ($\omega_z/\omega_m=0.60$, 0.80, 0.95).
Cooling results from different pulse numbers, $N=99$ and $N=199$, are compared. These results show, by employing the periodical $\sigma_z$ pulses, we can cool the resonator efficiently within a wide range of defect energy gaps.

\begin{figure}[htbp]
\begin{center}
\includegraphics[width=0.5\textwidth]{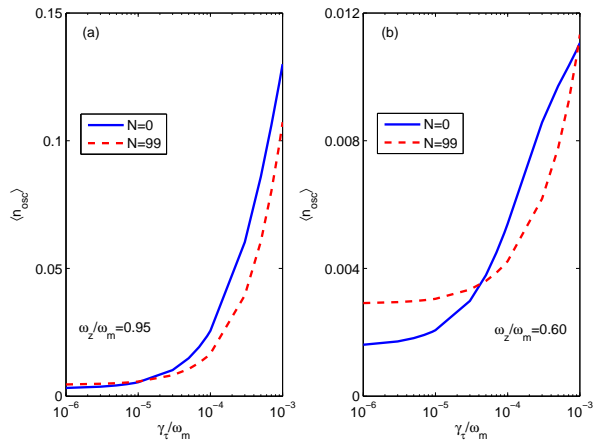}
\end{center}
\caption{\label{fig3}(Color online): Residual phonon number ($\langle n_{osc}\rangle$) versus $\gamma_\tau$. Different number of $\sigma_z$ pulses ($N$=0, 99) are applied. The value of $\langle n_{osc}\rangle$ is chosen at time $\omega_mt=200$, where the residual phonon number has been stable already. Parameters set: $\kappa/\omega_m=0.15$, $\gamma_m/\omega_m=10^{-6}$, $g/\omega_m=0.05$, $\lambda/\omega_m=0.05$, $\Delta_L/\omega_m=-1$, $T=0.1K$.}
\end{figure}

Finally, we compare the cooling performance of the resonator with different TLS damping rates $\gamma_\tau$ in Fig.~\ref{fig3}. The near resonance condition is chosen ($\delta\omega/\omega_m=0.05$), and the parameters satisfy, $\delta\omega\leq\lambda<\kappa$.
From Fig.~\ref{fig3}.~(a), we find that when the TLS damping rate is small enough (between $10^{-6}$ and $10^{-5}$), the cooling performance of the resonator is insensitive to the pulse number applied and the residual phonon number of the resonator is very small ($\simeq0.004$). In Fig.~\ref{fig3}.~(b), we change the defect energy gap to a new value, $\omega_z/\omega_m=0.60$, and the resonance condition is not satisfied, $\delta\omega>\lambda=0.05\omega_m$. When the TLS damping rate is larger than $5*10^{-5}$, the resonator is cooled efficiently with the pulse number $N=99$, while the result is not effective with no $\sigma_z$ pulses applied.

%When the TLS damping rate is larger than $10^{-5}$, compared with no incident $\sigma_z$-pulse case, we achieve a smaller residual phonon number of the resonator with $N=99$ incident $\sigma_z$-pulses. Taking $\gamma_\tau/\omega_m=10^{-3}$ as an example, the residual phonon number is 0.1072 for the case of $N=99$, when comparing with the case of no incident $\sigma_z$-pulse, a decrease of 17.5\% is obtained. So it is helpful to cool the resonator by injecting a periodical $\sigma_z$-pulses sequence.  the residual phonon number of the resonator with $N=99$ is smaller than that of no incident pulse case.

\section{Conclusion}
%In summary, we use the periodical $\sigma_z$-pulses to eliminate the bad effect from the defects in cooling the mechanical resonator. Compared with the no $\sigma_z$-pulse case, we can lower the phonon number of the resonator with different defect energy gaps and different TLS damping rates. In our discussion, the introduction of $\sigma_z$-pulse can remove the interaction between the TLS and the heat bath up to first order of time. The cooling of resonator is mainly from the radiative pressure of the cavity mode, and the achieved cooling performance is bounded. Considering the inevitable interaction between defects and resonator, we hope to take advantage of coexistence of defects and cavity mode, and cool the resonator more efficiently next.

In summary, we introduce periodical $\sigma_z$ pulses to eliminate the bad effect from the defects in cooling the mechanical resonator. The periodical $\sigma_z$ pulses can remove the interaction between the TLS and the heat bath up to the first order of time. By applying $\sigma_z$ pulses, we can cool the resonator efficiently with different defect energy gaps and different TLS damping rates. Other designed pulse sequences eliminating the interaction more than first order of time~\cite{Khodjasteh_PRL05, Uhrig_PRL07, Yang_PRL08, Uhrig_PRL09, West_PRL10} might be more efficient to cool the resonator. This deserves further study in the future.

%Compared with the no $\sigma_z$-pulse case, we can lower the phonon number of the resonator more efficiently

\section*{Acknowledgement}
 We thank W. J. Yang for helpful discussions. We acknowledge the financial support in part by the 10000-Plan of Shandong province, and the National High-Tech Program of China grant No. 2011AA010800 and 2011AA010803, NSFC grant No. 11174177 and 60725416.
{}
\end{document}